\documentclass[12pt,preprint]{aastex}
\usepackage{graphicx}  
\usepackage{lscape}
\usepackage{epsfig}
\usepackage{color}
\usepackage{url}
\newcommand{\psr}{PSR~B1259$-$63}
\newcommand{\sta}{LS~2883}
\newcommand{\binary}{PSR~B1259$-$63/LS~2883}
\newcommand{\gr}{$\gamma$-ray}
\newcommand{\grs}{$\gamma$-rays}

\def\etal{{\it et al.~}}

\begin{document}

\title{The hour-timescale GeV flares of PSR~B1259$-$63 in 2017}
  \author{P.~H.~T. Tam$^{1}$, X.-B. He$^{1}$, P.~S. Pal$^{1}$, Yudong Cui$^{1}$}
\affil {$^1$ School of Physics and Astronomy, Sun Yat-sen University, Zhuhai 519082, China\\
}
\email{tanbxuan@sysu.edu.cn, hexb7@mail2.sysu.edu.cn, parthasarathi.pal@gmail.com, cuiyd@mail.sysu.edu.cn}

\begin{abstract}
GeV flares from PSR~B1259$-$63/LS~2883 were seen starting around 30 days after the two periastron passages in 2010 and 2014. The flares are clearly delayed compared to the occurrence of the X-ray and TeV flux peaks during the post-periastron disk crossing. While several attempts have been put forward to explain this phenomenon, the origin of these GeV flares remains a puzzle. Here we present a detailed analysis of the observational data taken by the Fermi and Swift observatories over the 2017 September periastron passage. For the first time, we find short-lived but powerful GeV flares on time scales of down to three hours. The onset of the GeV flaring period in 2017 is also delayed compared to those seen in 2011 and 2014. Supplemented by a re-analysis of previous data, we compare the Fermi/LAT, Swift/XRT and Swift/UVOT light curves in 2017 with those taken over the 2010 and 2014 periastrons, and difference in UVOT light curves are noted.
\end{abstract}


\section{Introduction}

Gamma-ray binary systems are high-mass X-ray binaries (HMXBs) consisting of a neutron star or black hole in orbit with an O/B star. They provide a unique astrophysical laboratory to study the pulsar wind at sub-parsec scale through its interaction with the stellar wind. Currently, there are 7 known \gr~binaries (five listed in Dubus 2015, plus P3 in LMC, Corbet \etal 2016, and PSR J2032+4127 which has an orbital period of 45--50 years, Ho \etal 2016), but the PSR~B1259$-$63/LS~2883 system remains the only \gr~binary that (i) contains a pulsar, and (ii) emit repetitive orbital phase-dependent radio, X-ray and \gr~flares seen more than once. \psr~(with a spin period of 47.8~ms) orbits around the Be star LS~2883 with a period of 3.4 years in a highly eccentric orbit ($e\sim0.87$).  

\sta~\citep{1259_radio_94} hosts a stellar disk that is inclined with respect to the pulsar orbital plane, such that the pulsar crosses the disk twice around each periastron passage~\citep{Chernyakova_06}. The pulsar generates a powerful pulsar wind (PW) that interacts with the stellar wind/disk every time when the two bodies are close to each other, emitting non-thermal, unpulsed radiation, in radio~\citep[e.g.,][]{1259_radio_05}, X-rays~\citep[e.g.,][]{Chernyakova_09}, GeV \grs~\citep{Tam_1259_2015,Caliandro_15}, and TeV $\gamma$-rays~\citep{hess_1259_05,hess_1259_09,HESS_1259_2011}. The system was first detected in \grs~by H.E.S.S. shortly before and for up to 100 days after its 2004 periastron passage (March 7, 2004), making it the first variable Galactic TeV source~\citep{hess_1259_05}. Since the X-ray and TeV flux peaks occur near the two disk passages, it is customary to explain the bulk of X-rays and TeV \grs~as a result of pulsar wind-stellar disk shock acceleration.

With the launch of the Fermi Gamma-ray Observatory in 2008, GeV \grs~were first discovered from PSR~B1259$-$63/LS~2883 around the 2010 periastron passage. In particular, a GeV flare was first seen about 30 days past periastron. This GeV flare was not expected, and is still one of the major unresolved questions among \gr~binary studies~\citep{Dubus_review}. The GeV flare was found to repeat in 2014 at similar orbital phase, confirming the repetitive nature of the GeV flaring period. In 2014, contemporaneous X-ray activity during the GeV flares was found by multi-wavelength studies using NuSTAR, Swift, and Fermi data~\citep{Tam_1259_2015, Chernyakova_15}. The GeV flares do not have a corresponding counterpart in TeV~\citep{HESS_1259_2011}.

The radiation mechanisms of the broadband radiation from \binary, as well as the GeV flare seen in early 2011, are unclear. Electrons accelerated in the shock between the PW and stellar wind can produce synchrotron radiation and/or upscatter stellar or disk photons from \sta~to produce inverse-Compton (IC) radiation~\citep{Tavani97,Kirk99,Dubus06,Bogovalov08,Khangulyan11,Kong_model_11,Mochol13,Takata12}. The unshocked PW particles may also generate \grs~\citep{Khangulyan12}. The interaction between the stellar disk and the pulsar~\citep{Chernyakova_14}, as well as Doppler boosting~\citep{Dubus10,Kong_model_12} were also suggested to play a major role in producing the GeV flares. 

\psr~crossed the periastron in September 2017. In this paper, we present the Fermi's LAT, and Neil Gehrel Swift Observatory's XRT and UVOT analysis results of the binary \binary, in 2017 and in previous passages. The 2010, 2014 and 2017 periastron times are taken to be 2010 Dec 14 UT 16:39:02 (MJD 55544.69377),  2014 May 04 UT 10:02:21 (MJD 56781.418298) and  2017 Sep 22 UT 03:25:40 (MJD 58018.142824), based on the most updated orbital parameters as presented in~\cite{orbital_shannon_14}.

\section{Data analysis}
\subsection{Fermi}
\label{sect:fermi_analysis}
The Fermi/LAT data were analyzed using the Fermi Science Tools v10r0p5 package. For the 2010, 2014 and 2017 periastron passages, we used reprocessed Pass 8 data belonging to the SOURCE event class. We excluded events with zenith angle greater than 90$^{\circ}$ to reduce the photons from the Earth. The instrument response functions were "P8R2$\_$SOURCE$\_$V6" in our analysis. We performed a binned maximum-likelihood analysis (gtlike) of the data within a circular region of interest (ROI) with a radius 20$^{\circ}$ centered on the PSR B1259-63. We considered the background contribution of the Galactic diffuse model (gll$\_$iem$\_$v06.fits) and the isotropic background (iso$\_$P8R2$\_$SOURCE$\_$V6$\_$v06.txt), as well as known gamma-ray sources in the 3FGL catalog~\citep{3fgl_cat}. The spectral parameters of the sources further away than 5$^{\circ}$ from \psr~were fixed to the catalog values. For the sources within 5$^{\circ}$, the normalization is left free. The normalization factor of the two diffuse emission components were also allowed to vary. The spectrum of \psr~is assumed to be a power law:

\begin{equation}
\frac{dN}{dE} = N_0 \left(\frac{E}{E_0}\right)^{-\Gamma}.
\end{equation}

In all plots showing Fermi-LAT data, detections are defined as having a test-statistic (TS) value of 9 or above. When TS is smaller than 9, upper limits at the 90\% confidence-level are shown.

\subsection{Swift}
In 2017, a \textit{Swift} monitoring campaign consisting of 22 observations was performed, resulting in a total exposure of 29.5~ks spanning from September 17th (MJD 58013.7, $t_\mathrm{p}-4.3$d) to December 23rd (MJD 58110.3, $t_\mathrm{p}+92.3$d), as listed in details in Table~\ref{xrt_obs}.

For Swift-XRT data reduction, the level 2 cleaned event files of SWIFT-XRT 
are obtained from the events of photon counting (PC) and Window timing (WT) 
mode data with {\it xrtpipeline}. The spectra are extracted from a circular region in the best source position with 20 arcsec radius. The background is estimated from an annular region in the same position with radii from 30 to 60 arcsec. The ancillary response files (arfs) are extracted with {\it xrtmkarf}. The PC 
redistribution matrix file (rmf) version (v.12) is used in the spectral fits.
XRT-PC spectra are then analysed with XSPEC(v12.9.1m) with a \emph{ztbabs} $\times$ \emph{zpo} model. The column density ($n_\mathrm{H}$) is fixed within the range $5 - 7 \times 10^{22}$~cm$^{-2}$ (Tam et al., 2015). From fitted spectra, unabsorbed flux is calculated from 0.3-10 keV in cgs units for all 
observations. 

For Swift-UVOT data reduction, all extensions of sky images are stacked with {\it uvotimsum}.
The source magnitudes are derived with 3-$\sigma$ significance level from the circular region of 5 arcsec radius in the best source position of the stacked sky images from all the filters with {\it uvotsource}. The background is estimated from an annular region centered at the same position with an inner and outer radius of 10 arcsec and 20 arcsec, respectively.
 
\section{Results}
\subsection{Properties of the \gr~Light Curve in 2017}

\begin{figure}
\includegraphics[angle=270,scale=0.7]{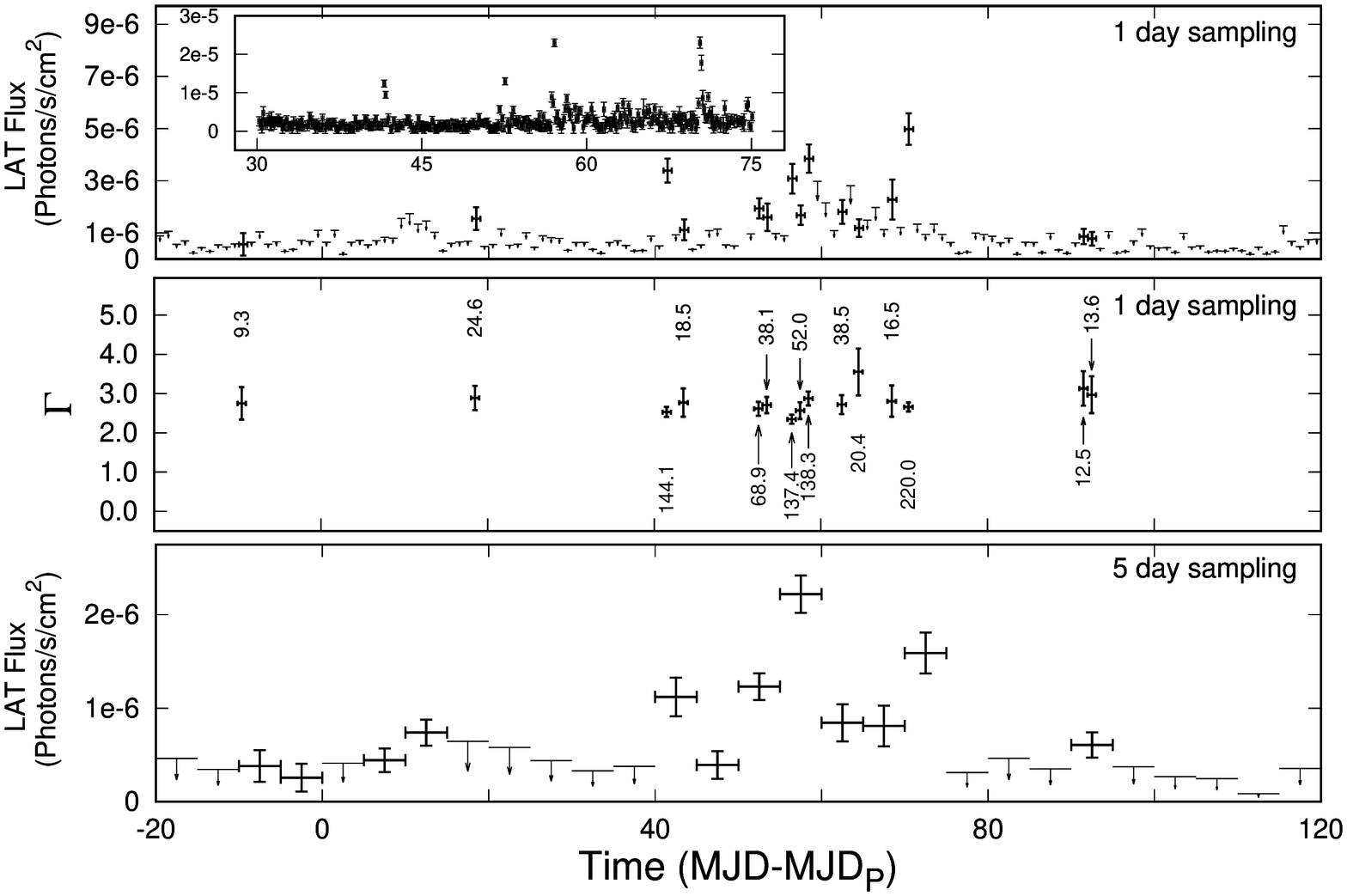}
\caption{Comparison of the LAT light curve over the 2017 periastron passage of PSR~B1259-63
with different sampling bins. The upper and lower panels were obtained using maximum likelihood analysis as described in Sect.~\ref{sect:fermi_analysis}, using 1-day and 5-day sampling, respectively. The inset of the upper panel was obtained using aperture photometry using 3-hour sampling, and therefore background is not subtracted. The TS values and the best-fit photon indices using 1-day sampling are shown in the middle panel.
}
\label{lc:2017}
\end{figure}

In Fig.~\ref{lc:2017} the LAT light curves of PSR~B1259-63 through the 2017 periastron passage with different data sampling (i.e., 3-hour, 1-day, and 5-day) are shown. The inset of the upper panel was obtained using aperture photometry with 3-hour sampling, in which 0.1--300~GeV photons were selected from a circle of radius 0.9 degree centered on \psr, therefore no background subtraction was done. The upper and lower panels were obtained using maximum likelihood analysis as described in Sect.~\ref{sect:fermi_analysis}, using 1-day and 5-day sampling, respectively. For 1-day time bins, the TS values and the best-fit photon indices from the likelihood analysis are shown in the middle panel.

Using a 5-day sampling, significant GeV emission was observed between $t_\mathrm{p}-$10~days to $t_\mathrm{p}+$15~days and  $t_\mathrm{p}+$40~days to  $t_\mathrm{p}+$75~days, and we regard the second period as the major GeV flaring period. However, as seen from the smaller time bins (i.e., 3-hour and 1-day), the GeV emission during this flaring period is highly variable over much shorter time scales, and several well-separated, sporadic major GeV flares can be seen. For example, the daily flux reaches 5$\times$10$^{-6}$~cm$^{-2}$~s$^{-1}$ (the highest daily flux, and daily TS value is 220) on the 71st day after periastron, and quickly drops below the 1-day sensitivity of the Fermi/LAT on the following day (with a TS value of 0), resulting in an upper limit of 1.3$\times$10$^{-6}$~cm$^{-2}$~s$^{-1}$. At four occasions, the daily photon flux is $\ga$3$\times$10$^{-6}$~cm$^{-2}$~s$^{-1}$ (which occur on 42, 57, 59, and 71 days after periastron, and with TS values all above 130 during these four days). After that, the GeV emission is no longer detected, apart from a short episode in $t_\mathrm{p}+$92--93 days. 

We show in Fig.~\ref{lc:2017} the evolution of the photon index in daily time scale. In contrast to the photon flux, the photon index does not change significantly over time, and remains at about 3.0.

Using a 3-hour sampling, it can be seen that the daily variation comes from shorter-time variation down to a few hours. The flux change can be up to more than one order of magnitude within 3 hours. This is the first time such very fast and large variation is ever reported for \binary. 

\subsection{Properties of the \gr~Light Curve in 2017 compared to those in 2011 and 2014}

\begin{figure}
\includegraphics[angle=270,scale=0.7]{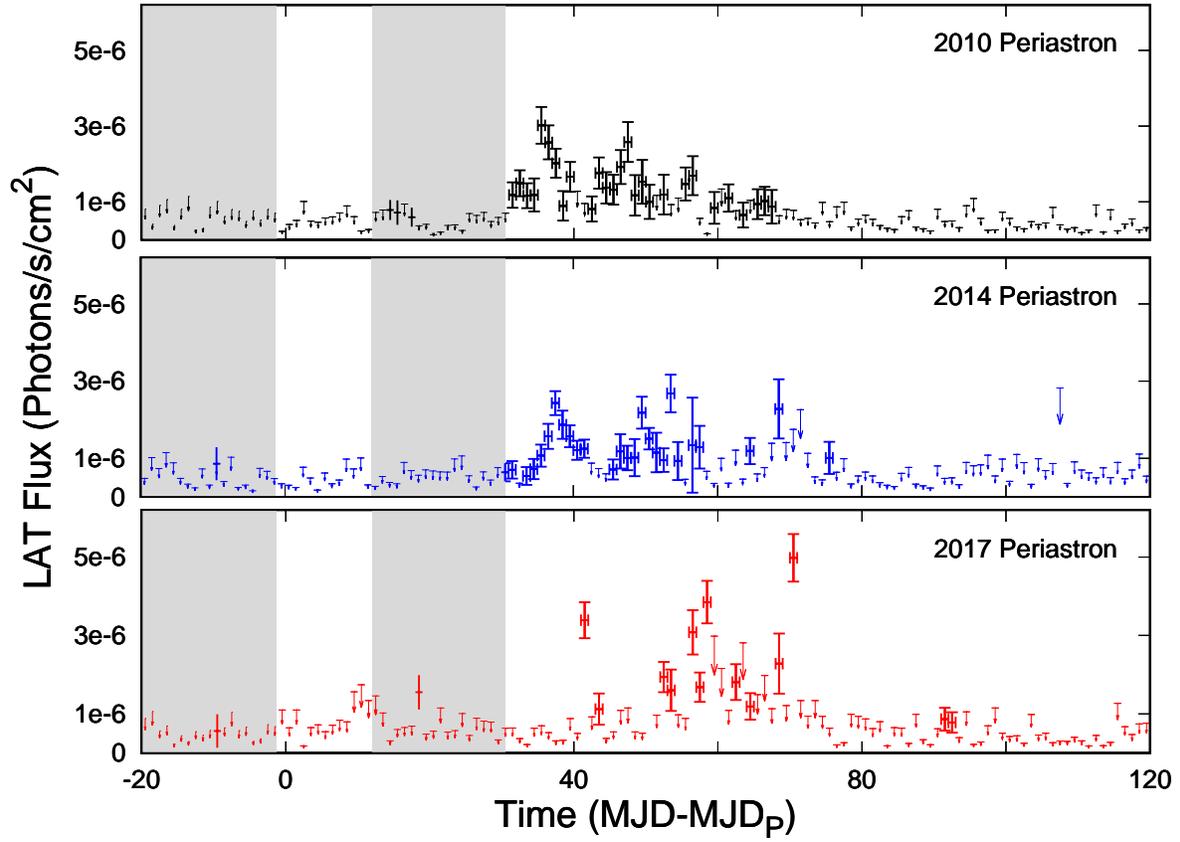}
\caption{Comparison among the LAT observations for 2010, 2014 and 2017 
periastrons of PSR~B1259-63 with 1 day sampling. The shaded region shows the FWHM of the Gaussian fits of the two X-ray flares in 2017, representing the stellar disk passage of \psr.
\label{1day_compare}}
\end{figure}

Fig.~\ref{1day_compare} shows the LAT daily flux variation of PSR~B1259-63 during and after the three periastrons. The top panel shows the flux variation during the 2010 periastron with black points. The middle panel shows the flux variation during 2014 periastron with blue points. The bottom panel shows the flux variation during 2017 periastron with red points. For all 3 panels the shaded regions indicate the stellar disk passage of \psr, which we inferred from the X-ray data taken over the 2017 periastron (see Sect.~\ref{Sect:x_gamma}).

It is clear that all GeV major flares occur at orbital phase well after the second disc crossing, and the 2017 flares start at orbital phase clearly different from those in 2011 and 2014. The light curve also differs, e.g., in 2017 the flares are more short-lived, while the GeV flares in 2011 and 2014 are more slowly-varying, including the onset of the flares at $\sim t_\mathrm{p}+$30--40 days.

\begin{figure}
\includegraphics[angle=270,scale=0.35]{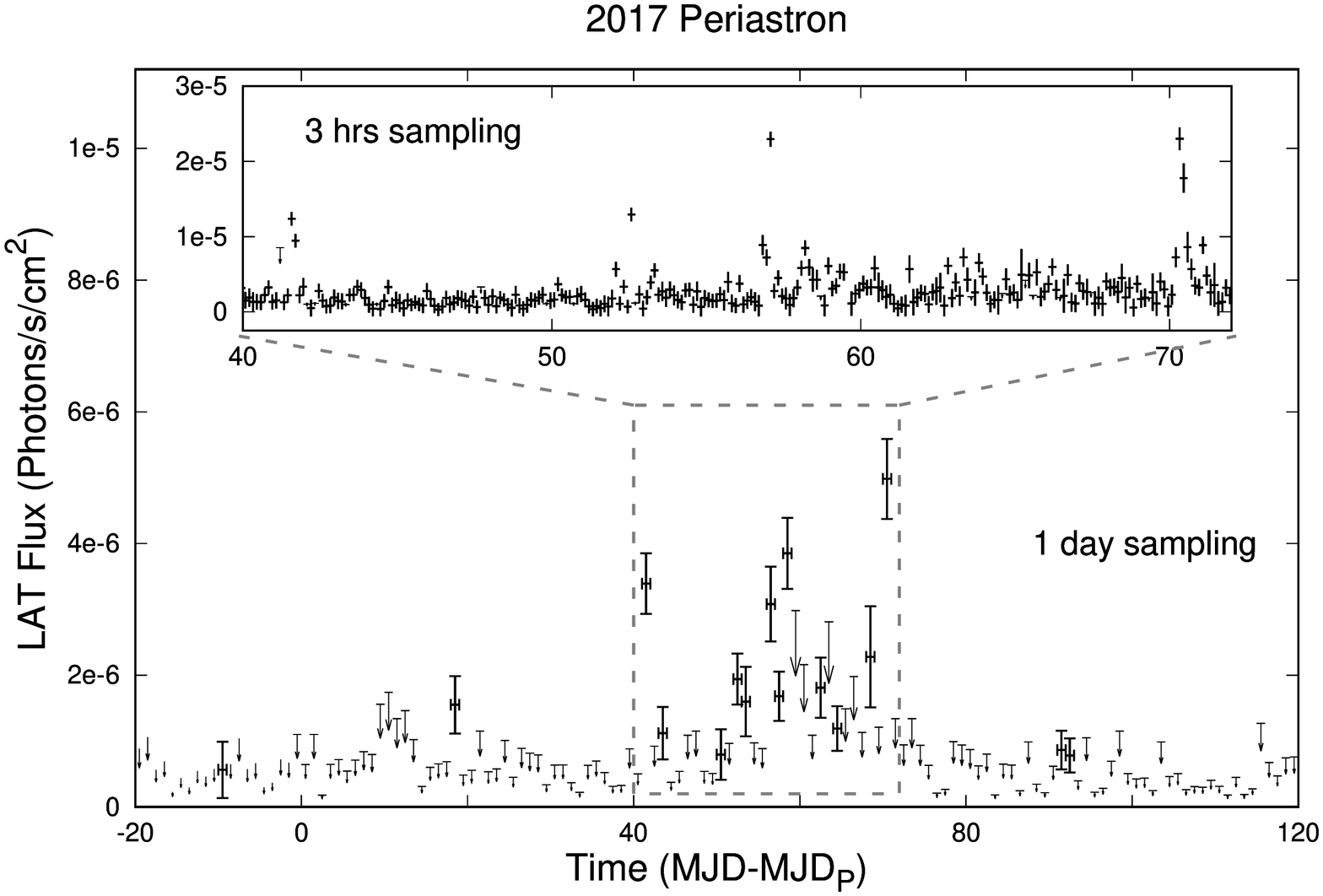}
\includegraphics[angle=270,scale=0.35]{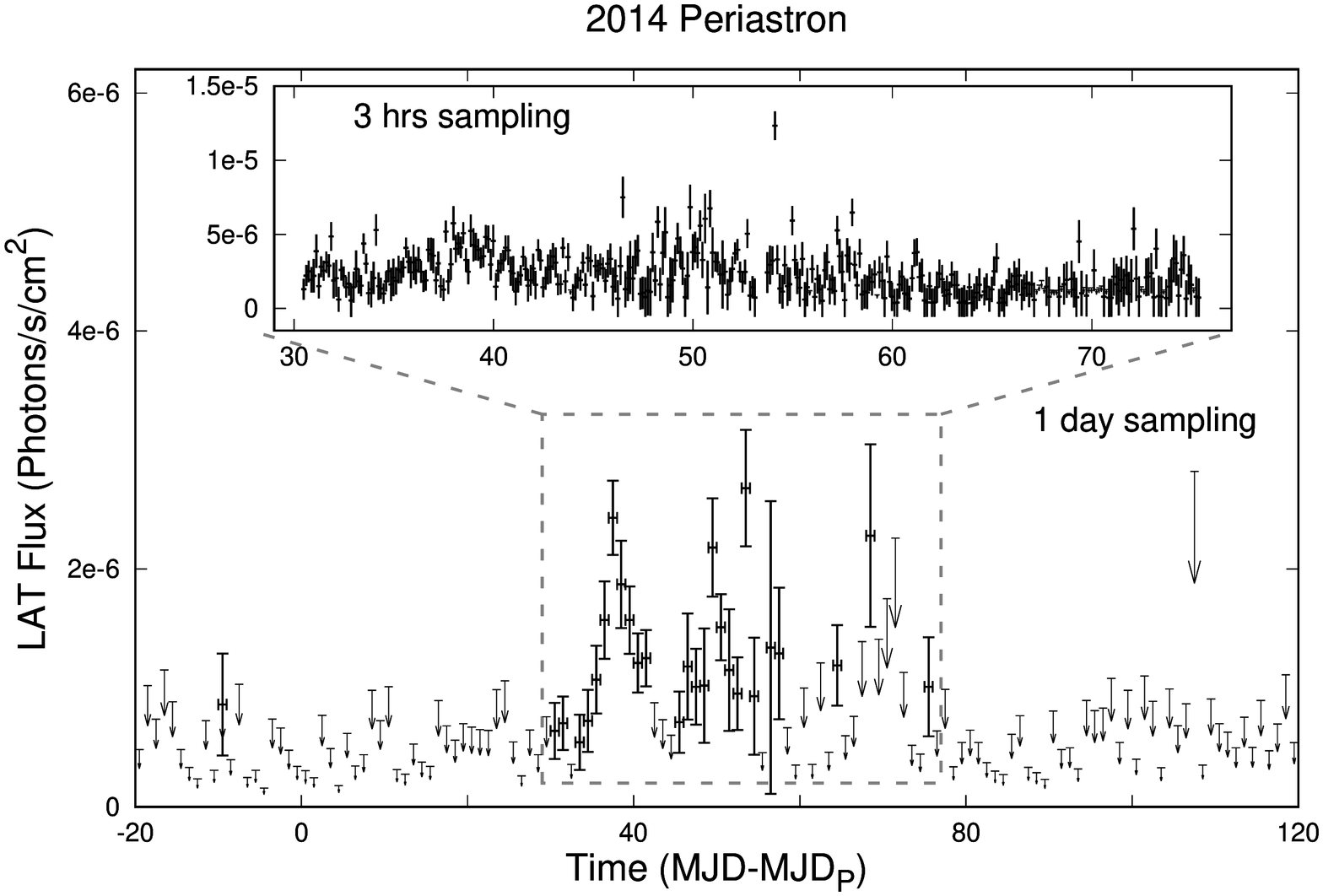}
\includegraphics[angle=270,scale=0.35]{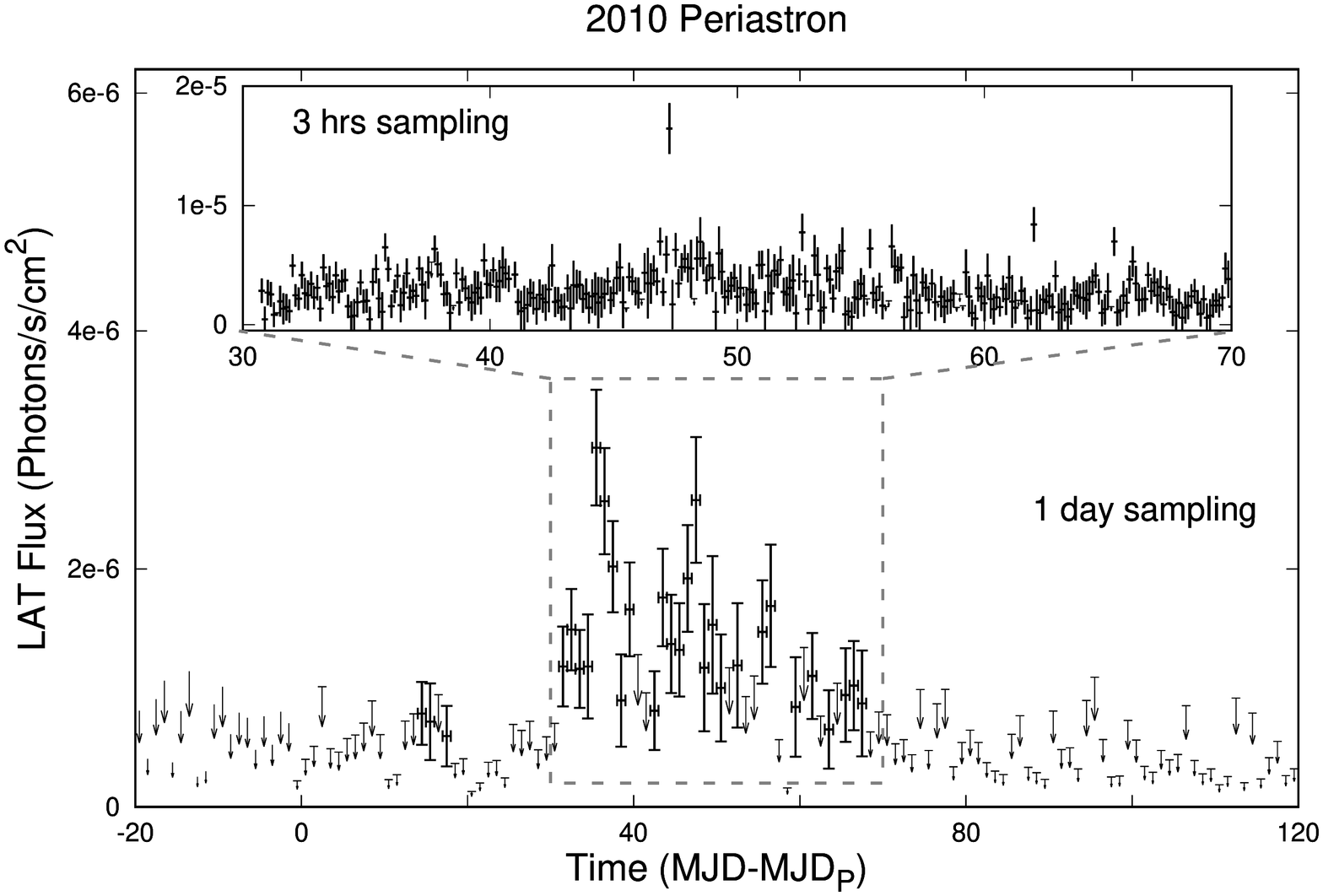}
\caption{
First panel: LAT flux variation of 2017 periastron with 1 day sampling. 
In the inset 3 hrs sampling of flux variation is shown from 40 to 70 days past periastron. 
Second panel: LAT flux variation of 2014 periastron with 1 day sampling. 
In the inset 3 hrs sampling of flux variation is shown from 30 to 75 days past periastron. 
Third panel: LAT flux variation of 2010 periastron with 1 day sampling. 
In the inset 3 hrs sampling of flux variation is shown from 30 to 70 days past periastron. 
\label{3h_compare}}
\end{figure}

To check whether the 2017 three-hour time scale flares are due to different data sampling, we also present 3-hour light curves in 2011 and 2014, again using aperture photometry, in Fig.~\ref{3h_compare}. One can see that the general flux variation over the two previous periastrons does not occur on such short time scales. Furthermore, even though some $\sim$3-hour timescale flares are also seen in 2011 and 2014 as well, the peak flux is the highest in 2017. This demonstrates that the difference in the appearance of the flares as seen in the year 2017 (as compared to 2011 and 2014) is robust against different time bins used.

\subsection{Characterising the GeV emission before and during the periastron}
\begin{figure}
\centering
\includegraphics[width=12cm, angle=0]{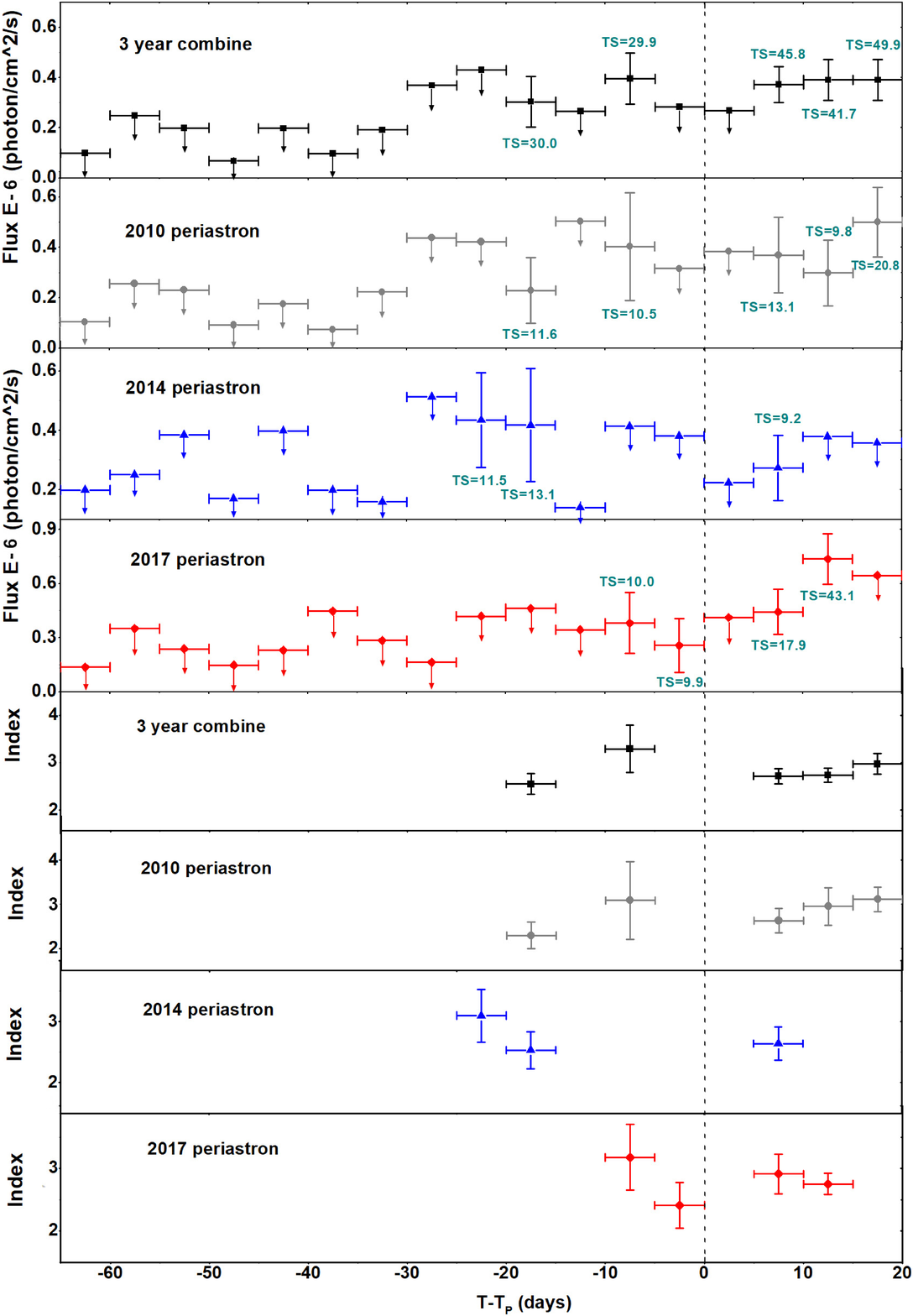}
\caption{Top four panels: the 100~MeV--300~GeV light curve of PSR B1259-63 close to the periastron passage, with 5-day binning. The TS values of those periods with significant detections (TS$>$9) are shown. Those data without a TS value next to it are upper limits. Bottom four panels: the photon index are only shown for periods with TS$>$9. }
 \label{pre-periastronGeV}
\end{figure}

With three periastron data, we attempt to characterise the GeV emission before and during the periastron. Data for each periastron as well as combing data in the three passages are shown in Fig.~\ref{pre-periastronGeV}. Since there is no evidence of short flares, 5-day sampling is employed to increase the photon statistics. Fig.~\ref{pre-periastronGeV} shows the GeV emission before, during, and shortly after each periastron. 

With GeV data taken during three periastrons, it is now established that there is clearly a GeV enhancement or brightening from about $t_\mathrm{p}-$20~days to $t_\mathrm{p}+$20~days (when stacked data from the three passages are used), which is well separated in time from the post-periastron GeV flare emission.

\subsection{The X-ray and \gr~light curves in 2014 and 2017}
\label{Sect:x_gamma}
\begin{figure}
\includegraphics[angle=270,scale=0.35]{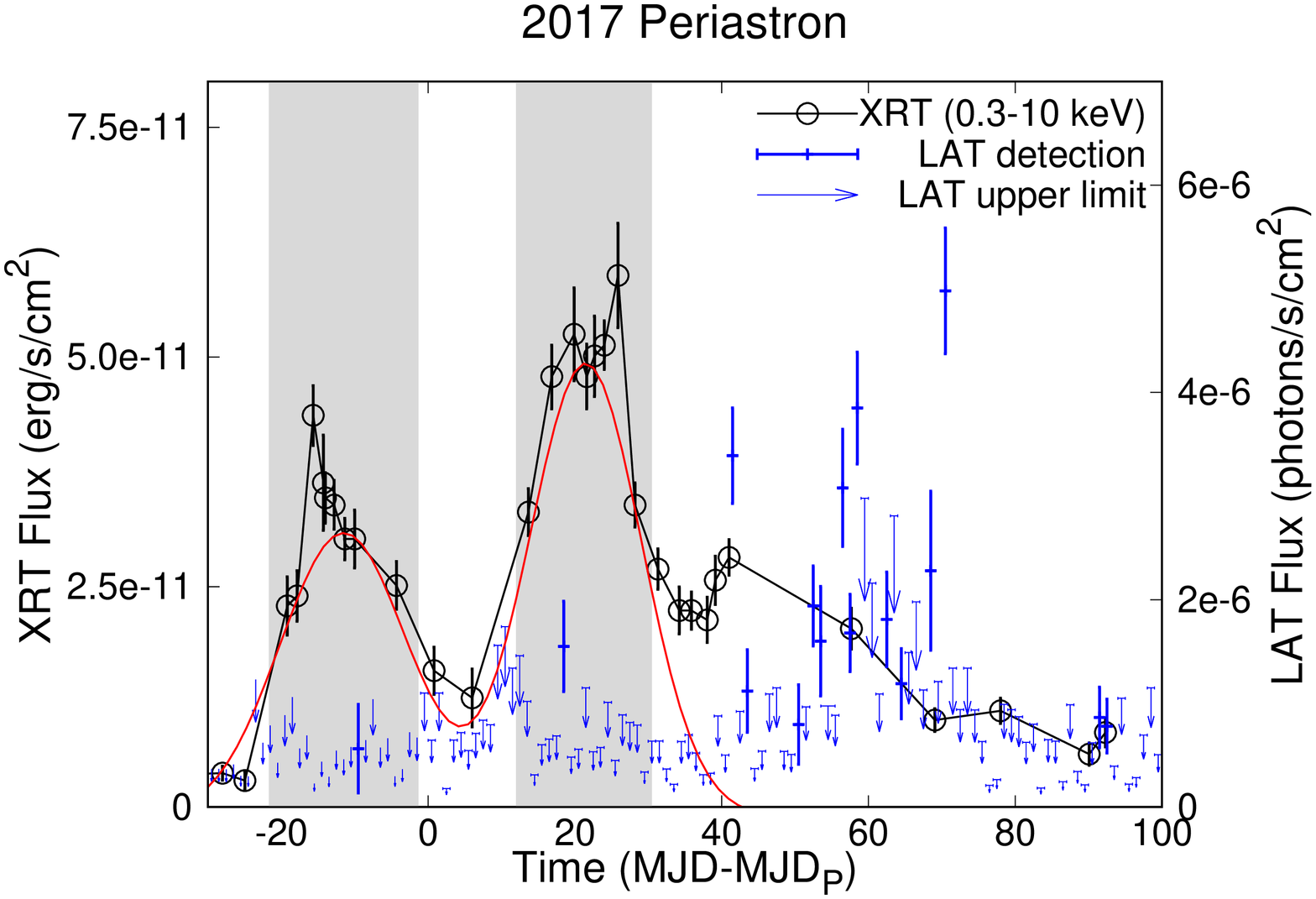}
\includegraphics[angle=270,scale=0.35]{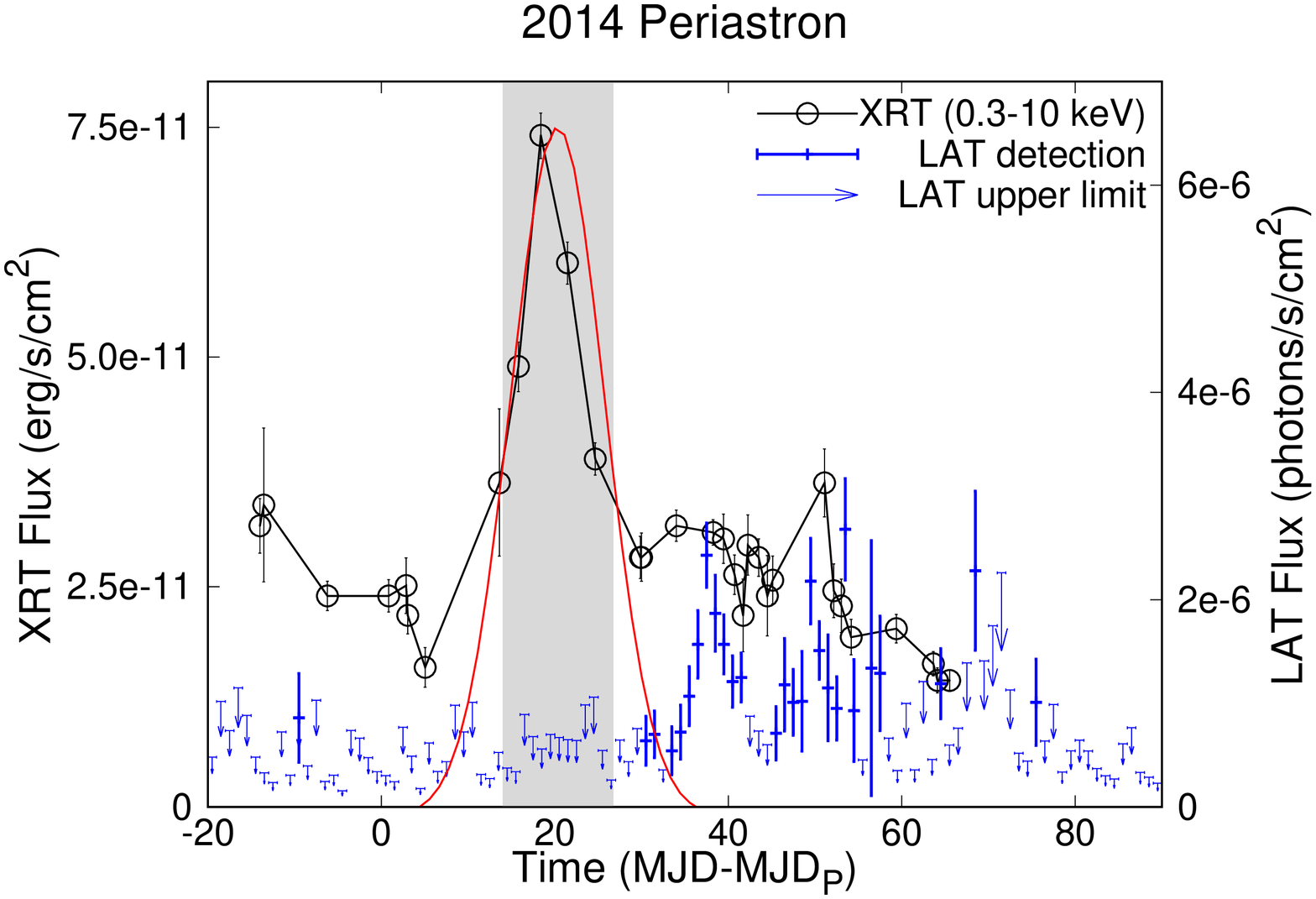}
\caption{Left panel: Comparison between XRT (black) and LAT (blue) observations (1-day sampling) of the 2017 periastron. The Gaussian fits (in red) return two X-ray peaks which are centered at $t_\mathrm{p}-$11.5 days and $t_\mathrm{p}+$21.5 days and a FWHM width of 10.5 days each. Right panel: Comparison between XRT (black) and LAT (blue) observations (1-day sampling) of the 2014 periastron. In X-rays, only the post-periastron peak is evident and the Gaussian fit is centered at $t_\mathrm{p}+$20.5 days and has a FWHM width of 6.5 days.
\label{xl}}
\end{figure}

In X-rays, the 2017 light curve (also presented in Table~\ref{xrt_results}) is similar to the 2014 one (Fig.~\ref{xl}), and the slight difference could be caused by the inhomogeneous Be disk or the disk evolution. Since the X-ray light curve is best sampled in 2017, we obtain the Be disk crossing time of the pulsar by the two X-ray broad peaks in 2017 (using Gaussian fits), which is well explained by the terminal shock model of the pulsar wind~\cite{Chernyakova_06}. The Gaussian fits return two peaks which are centered at $t_\mathrm{p}-$11.5 days and $t_\mathrm{p}+$21.5 days and a FWHM width of 10.5 days each. We suggest that the first disk crossing is from $t_\mathrm{p}-$22 to $t_\mathrm{p}-$1 days, and the second crossing is from $t_\mathrm{p}+$12 to $t_\mathrm{p}+$31 days based on the X-ray data obtained in 2017. In 2014, only the post-periastron peak is evident and it is centered at $t_\mathrm{p}+$20.5 days and a FWHM width of 6.5 days.

In Fig.~\ref{xl} X-ray and $\gamma$-ray flux variation is compared over 2017 and 2014 periastrons. While in 2014 there is correlated X-ray/\gr~ activities during the GeV flares (c.f., \cite{Tam_1259_2015,Chernyakova_15}), no correlation is visible in 2017. This may, however, be due to much sparse data coverage during the GeV flares by Swift in 2017.

\subsection{UVOT results of the last three periastrons}

\begin{figure}
\includegraphics[angle=270,scale=0.35]{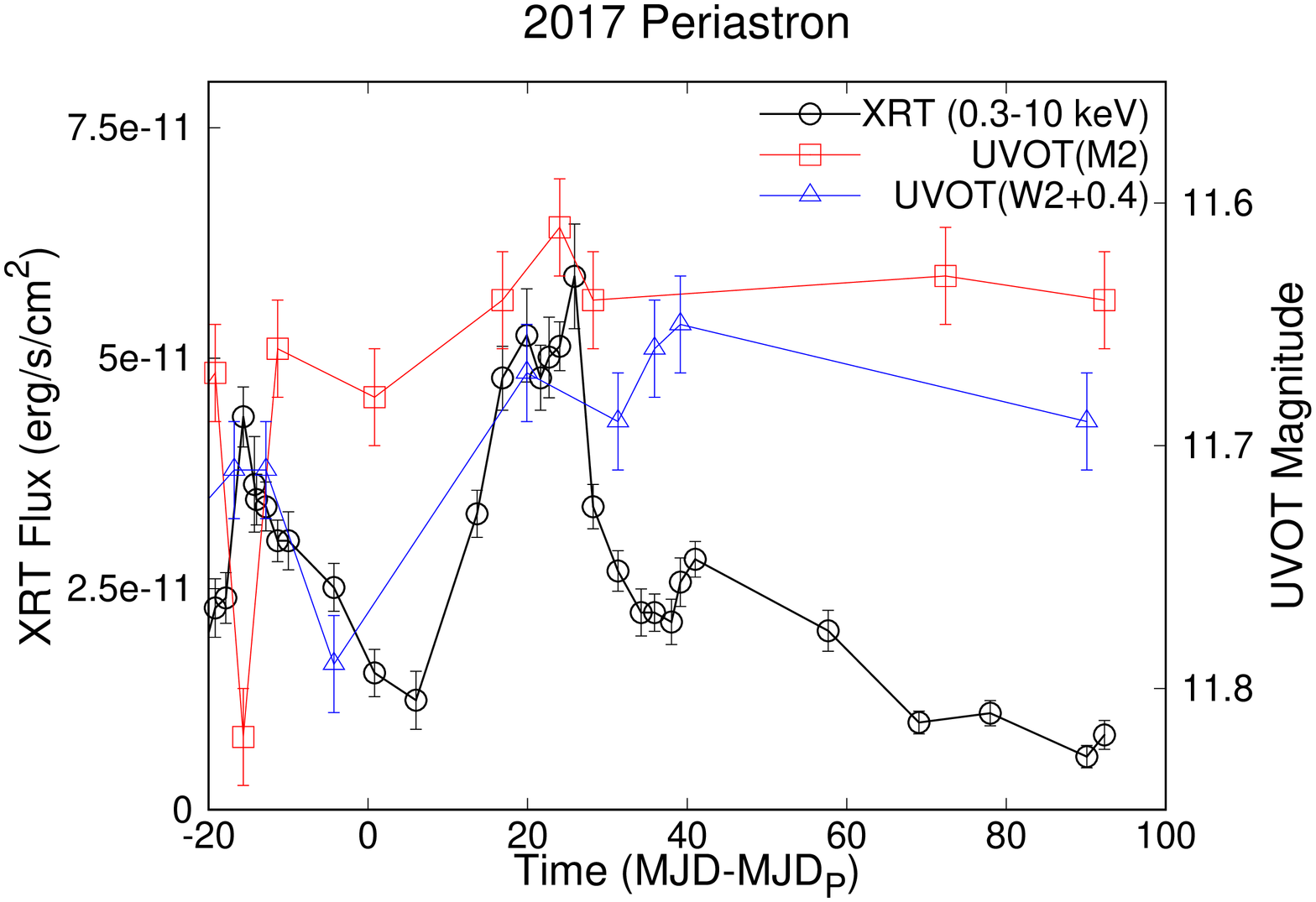}
\includegraphics[angle=270,scale=0.35]{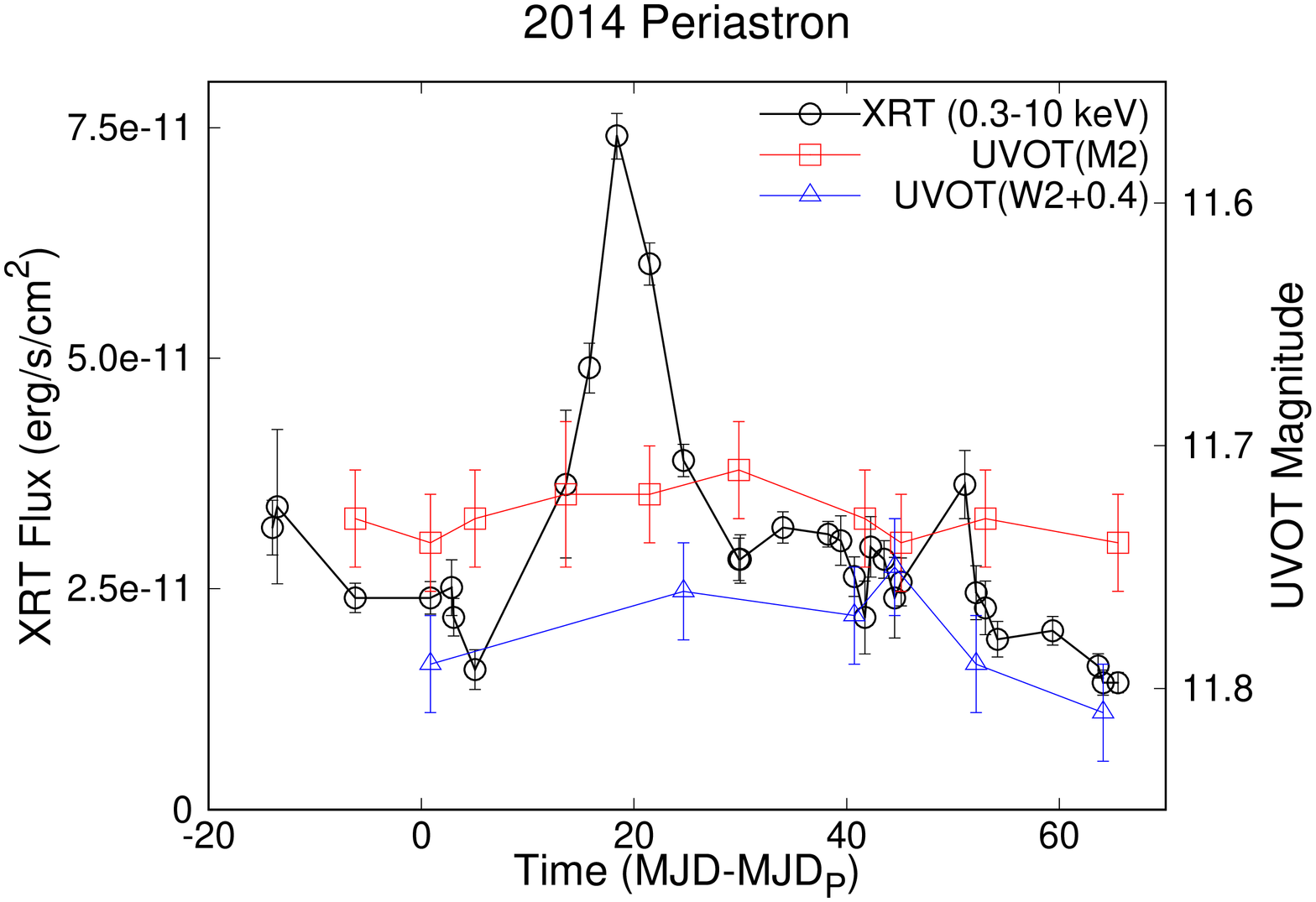}
\includegraphics[angle=270,scale=0.35]{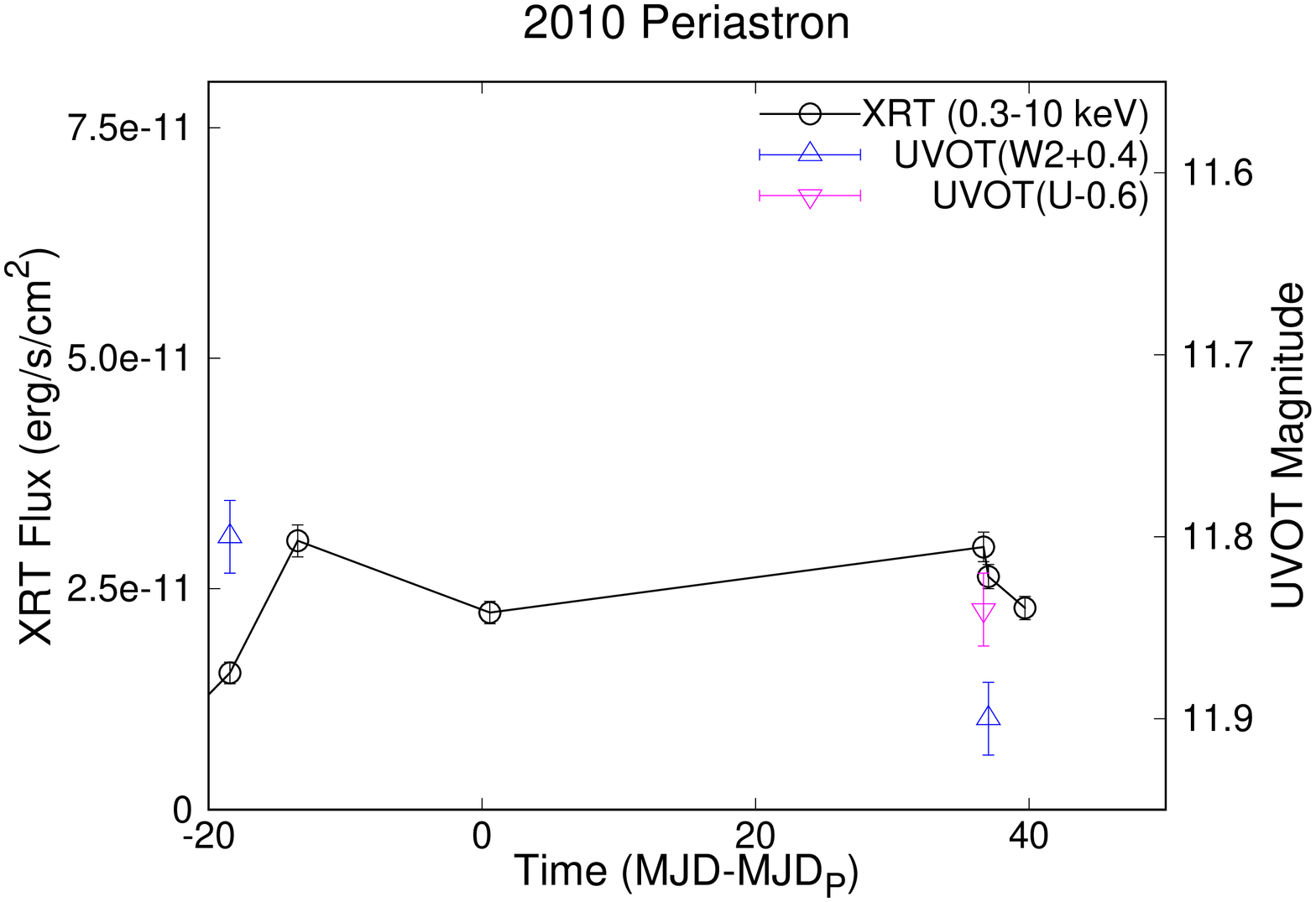}
\caption{Comparison between XRT and UVOT observations of the 2010, 2014, and 2017 periastrons. The W2 filter magnitude is offset by $+$0.4, and the U filter by $-$0.6.
\label{xu}}
\end{figure}

In Fig.~\ref{xu} the X-ray and UV/optical variation is compared during the three periastron passages. There is no significant change in the shown magnitudes for the 2014 periastron. In 2017, the situation is rather different: 1) there is significant evolution in the W2 magnitude, as well as the M2 magnitude; 2) the W2 magnitude seems to follow the trend seen by XRT, which is observed for the first time. These changes may be related to the two disc crossings as the UVOT flux changes is most visible before or after the crossings of the disc.

\section{Discussion}
\label{compare}
The most striking feature from this work is the different temporal evolution of the GeV flare occurring in 2017 compared to that seen in 2011 and 2014.

Previous Fermi analysis works by \cite{Tam_1259_2015}, \cite{Chernyakova_15}, and \cite{Caliandro_15} have argued that the GeV light curves in 2010 and 2014 seems to show a very similar onset times of the GeV flares $\sim$30~days past the periastron passage, and to a lesser extent, the flare evolution up to $t_\mathrm{p}+$75 days. However, during the most recent 2017 periastron passage, our Fermi analysis work has discovered very short yet very powerful GeV flares on time scale down to 3 hours in 2017, which is limited by the cadence of the Fermi observations of the source in the all-sky observing mode. In sharp contrast to the light curve, the GeV spectrum remains rather stable both over each periastron passage, as well as between flares that are separated by 3.4 years.

So far, the GeV flaring period lasting for 1--2 months which are still periodic (i.e., occurring after the second disk crossing) suggest that the GeV flares are likely associated with the pulsar passing through the dense gas in the Be star disk. Meanwhile, 
the random short flares especially seen in 2017 require a source other than the accelerated 
particles in the terminal shock. This terminal shock model has explained the periodic 
light curve of the non-thermal X-ray emission very well, see e.g.~\cite{Takata12}. 
In previous modeling work on the GeV flare, \cite{Khangulyan12} used the 
inverse Compton (IC) scattering of the unshocked pulsar wind with soft photons from the
 Be star, while \cite{Dubus13} brought up an IC model using the X-ray synchrotron radiation from 
the shock as the IC target photons. \cite{Kong_model_12} used the synchrotron model with 
relativistic flows beaming towards Earth along the bow shock tails.
All of these previous models, which are heavily depending on the terminal shock geometry, 
lack of explanation for either the delay of the GeV flaring period or the sporadic hour-timescale flares. 

Noticeably, to explain the GeV flares in Crab Nebula with time scales of several days,
 \cite{Lyutikov2012} has introduced the corrugation perturbations of the
 terminal shock, which allows fast changes in the Doppler beaming of the post-shock synchrotron
emission, and the GeV time scale of such model relies on overall dynamics of the termination shock, 
rather than on the decaying time of the high energy particles. However, in such case, 
we can only see GeV flares when the terminal shock is formed on the observer's side. 
As another attempt to explain the fast GeV flares in PSRs, magnetic reconnection is 
also a plausible choice, see e.g. \cite{Cerutti2014}.

Recently, \cite{Yi2017} raised the idea that the delay of the entire GeV flare 
package could be explained by the time scale of forming an accretion disk. This work has used the accretion disk to gain sufficient soft photons for the IC process of the pulsar wind, yet it did not provide an explanation for the very short flares. GeV emission regions with length scale down to the size of the accretion disk or even down to the NS itself remain possible candidates in explaining the $<$3 hour GeV flares. So far, there is no direct observational evidences of an accretion disk or a periodic NS glitch correspond to each periastron passage~\citep{Yi2018}.

Apart from the GeV flares, in this work we also establish the GeV brightening from about $t_\mathrm{p}-$20~days to $t_\mathrm{p}+$20~days which is well separated in time from the flares. The origin and emission mechanism of this brightening is unclear, and it would be interesting to see if it is due to either the close encounter between the pulsar and the massive star~\citep{Tavani97,Kirk99}, or the pre-periastron disk passage~\citep{Yi2017}.

To conclude, we found short-lived but powerful GeV flares on time scales of down to about three hours, after the recent 2017 periastron passage. The onset of the GeV flaring period in 2017 is also delayed compared to those seen in 2011 and 2014. On the other hand, the X-ray flux show similar variation from 2010 to 2017. During the 2017 passage, the UV flux shows significant variation in two filters and it may be correlated with the X-ray flux variation. Observing \binary~in next periastron passages will help us to get more ideas about the hidden physics and emission mechanisms that produce the GeV flares.

\acknowledgments
We thank KS Cheng, CW Ng, KL Li, J Takata, and AKH Kong for valuable discussion. We acknowledge the use of data and software facilities from the FSSC, managed by the HEASARC at the Goddard Space Flight Center. This work is supported by National Natural Science Foundation of China (NSFC) through grants 11633007, 11661161010, and U1731136.

\clearpage

\begin{table}
\begin{center}
\begin{tabular}{ccccccc}
\hline
Obs-Id & Date & Date(UTC) & t-$t_p$ & Data & Exposure & XRT-Rate \\
 & & MJD & Day & Mode & Sec & cnt/s \\
\hline
00030966071 & 2017-09-17 & 58013.70 & -4.30 & PC & 1271 & 0.25$\pm$0.01 \\
00030966072 & 2017-09-22 & 58018.82 & 0.82  & PC & 671.8 & 0.21$\pm$0.02 \\
00030966073 & 2017-09-28 & 58024.00 & 6.00  & PC & 393.4 & 0.17$\pm$0.02 \\
00030966074 & 2017-10-05 & 58031.67 & 13.67 & WT & 1487 & 2.12$\pm$0.04 \\
00030966075 & 2017-10-08 & 58034.85 & 16.85 & PC & 1388 & 0.49$\pm$0.02 \\
00030966076 & 2017-10-11 & 58037.90 & 19.90 & PC & 624.3 & 0.63$\pm$0.03 \\
00030966077 & 2017-10-13 & 58039.62 & 21.62 & PC & 1391 & 0.47$\pm$0.02 \\
00030966078 & 2017-10-14 & 58040.69 & 22.69 & PC & 870.6 & 0.50$\pm$0.02 \\
00030966079 & 2017-10-16 & 58042.01 & 24.01 & PC & 1988 & 0.63$\pm$0.02 \\
00030966080 & 2017-10-17 & 58043.87 & 25.87 & PC & 536.9 & 0.72$\pm$0.04 \\
00030966081 & 2017-10-20 & 58046.20 & 28.20 & PC & 1688 & 0.41$\pm$0.02 \\
00030966082 & 2017-10-23 & 58049.34 & 31.34 & PC & 1563 & 0.34$\pm$0.02 \\
00030966083 & 2017-10-26 & 58052.24 & 34.24 & PC & 1214 & 0.23$\pm$0.01 \\
00030966084 & 2017-10-27 & 58053.90 & 35.90 & PC & 1915 & 0.23$\pm$0.01 \\
00030966085 & 2017-10-30 & 58056.03 & 38.03 & PC & 1016 & 0.25$\pm$0.02 \\
00030966086 & 2017-10-31 & 58057.15 & 39.15 & PC & 1266 & 0.26$\pm$0.02 \\
00030966087 & 2017-11-02 & 58059.01 & 41.01 & PC & 2165 & 0.33$\pm$0.01 \\
00030966089 & 2017-11-18 & 58075.68 & 57.68 & PC & 1416 & 0.21$\pm$0.01 \\
00030966090 & 2017-11-30 & 58087.05 & 69.05 & PC & 2043 & 0.12$\pm$0.01 \\
00030966092 & 2017-12-09 & 58096.01 & 78.01 & PC & 2078 & 0.13$\pm$0.01 \\
00030966095 & 2017-12-21 & 58108.10 & 90.10 & PC & 1316 & 0.091$\pm$0.008 \\
00030966096 & 2017-12-23 & 58110.31 & 92.31 & PC & 1244 & 0.095$\pm$0.009 \\
\hline
\end{tabular}
\end{center}
\caption{Swift-XRT observation through the 2017 periastron of PSR~B1259-63. \label{xrt_obs} }
\end{table}

\begin{table}
\begin{center}
\begin{tabular}{ccccc}
\hline
t-$t_p$ & $n_H$ & $\Gamma$ & XRT-Flux & $\chi^2$\\
Day & $10^{22} cm^{-2}$ &  & $\times 10^{-11} erg/sec/cm^2$ & (dof) \\
\hline
\hline
-4.30 & 0.69 & $2.19~_{-0.35}^{+0.24}$ & $2.51~_{-0.26}^{+0.26}$ & 0.75(49) \\
0.82 & 0.5 & $1.66~_{-0.40}^{+0.48}$ & $1.58~_{-0.25}^{+0.26}$ & 0.89(21) \\
6.00 & 0.5 & $1.77~_{-0.67}^{+0.99}$ & $1.29~_{-0.32}^{+0.31}$ & 1.23(8) \\
13.67 & 0.5 & $1.58~_{-0.16}^{+0.18}$ & $3.31~_{-0.26}^{+0.26}$ & 0.84(81) \\
16.85 & 0.6 & $1.69~_{-0.30}^{+0.32}$ & $4.79~_{-0.35}^{+0.34}$ & 0.97(104) \\
19.90 & 0.68 & $1.51~_{-0.30}^{+0.22}$ & $5.25~_{-0.51}^{+0.50}$ & 0.81(62) \\
21.62 & 0.7 & $1.80~_{-0.22}^{+0.16}$ & $4.79~_{-0.35}^{+0.35}$ & 0.86(99) \\
22.69 & 0.7 & $1.82~_{-0.31}^{+0.19}$ & $5.01~_{-0.44}^{+0.44}$ & 0.70(70) \\
24.01 & 0.5 & $1.66~_{-0.11}^{+0.21}$ & $5.13~_{-0.27}^{+0.27}$ & 1.04(175) \\
25.87 & 0.6 & $1.42~_{-0.25}^{+0.28}$ & $5.89~_{-0.57}^{+0.57}$ & 0.82(61) \\
28.20 & 0.5 & $1.43~_{-0.14}^{+0.24}$ & $3.39~_{-0.24}^{+0.24}$ & 0.79(105) \\
31.34 & 0.57 & $1.52~_{-0.19}^{+0.24}$ & $2.69~_{-0.22}^{+0.22}$ & 0.80(85) \\
34.24 & 0.5 & $1.43~_{-0.27}^{+0.33}$ & $2.24~_{-0.25}^{+0.25}$ & 0.93(46) \\
35.90 & 0.7 & $1.63~_{-0.18}^{+0.17}$ & $2.24~_{-0.20}^{+0.20}$ & 1.0(74) \\
38.03 & 0.5 & $1.41~_{-0.24}^{+0.32}$ & $2.14~_{-0.25}^{+0.25}$ & 0.77(43) \\
39.15 & 0.5 & $1.63~_{-0.20}^{+0.34}$ & $2.57~_{-0.26}^{+0.26}$ & 0.64(55) \\
41.01 & 0.7 & $1.54~_{-0.18}^{+0.14}$ & $2.82~_{-0.20}^{+0.19}$ & 0.86(113) \\
57.68 & 0.5 & $1.39~_{-0.22}^{+0.35}$ & $2.04~_{-0.22}^{+0.22}$ & 0.7(47) \\
69.05 & 0.5 & $1.47~_{-0.25}^{+0.35}$ & $1.05~_{-0.13}^{+0.12}$ & 0.7(40) \\
78.01 & 0.6 & $1.64~_{-0.29}^{+0.33}$ & $1.15~_{-0.14}^{+0.13}$ & 1.1(43) \\
90.10 & 0.5 & $1.65~_{-0.42}^{+0.64}$ & $0.68~_{-0.13}^{+0.12}$ & 0.9(17) \\
92.31 & 0.7 & $1.44~_{-0.66}^{+0.38}$ & $0.91~_{-0.15}^{+0.16}$ & 0.8(18) \\
\hline
\end{tabular}
\end{center}
\caption{Spectral analysis result of the Swift-XRT data taken in late 2017. $n_\mathrm{H}$ is interstellar absorption obtained for the TBabs model. 
$\Gamma$ is the power-law index of the XRT spectrum. Unabsorbed flux is calculated for the energy range 0.3--10 keV. \label{xrt_results}}
\end{table}

\end{document}